\documentclass[11pt]{article}
\usepackage{epsf,afterpage,amsmath,amssymb}

\newcommand{\beq}{\begin{equation}}
\newcommand{\eeq}{\end{equation}\noindent}
\newcommand{\bear}{\begin{eqnarray}}
\newcommand{\eear}{\end{eqnarray}\noindent}

\renewcommand{\descriptionlabel}[1]%
{\hspace{\labelsep}\textsf{#1}}

\newenvironment{mydescription}[1]%
 {\begin{list}{}{%
  \setlength{\leftmargin}{10pt}%
  \setlength{\itemindent}{10pt}%
  \setlength{\rightmargin}{10pt}}}
  {\end{list}}

\begin{document}

\begin{flushright}
BI/TH-98/05 \\
June 1998
\end{flushright}

\begin{center}
\vspace{24pt}

{\LARGE \bf Three-Dimensional Simplicial Gravity and \\
   \vspace{5pt}  Degenerate Triangulations} 
\vspace{24pt}

{\Large \sl Gudmar Thorleifsson}
\vspace{12pt}

Fakult\"{a}t f\"{u}r Physik, Universit\"{a}t Bielefeld, 
 33501 Bielefeld, Germany
\vspace{12pt}

\begin{abstract}
I define a model of three-dimensional simplicial gravity using
an extended ensemble of triangulations where, in addition 
to the usual {\it combinatorial} triangulations, I
allow {\it degenerate} triangulations, i.e.\ triangulations
with {\it distinct} simplexes defined by the same set
of vertexes.  I demonstrate, using numerical simulations,
that allowing this type of degeneracy substantially reduces the 
geometric finite-size effects, especially 
in the crumpled phase of the model, in other
respect the phase structure of the model is not affected.   
\end{abstract}
\end{center}
\vspace{15pt}

\section{Introduction}

Among the different attempts to construct a viable theory
of quantum gravity, Euclidean quantum gravity is one of
few that allows for non-perturbative studies using
numerical methods.  In the Euclidean approach one works
with a functional integral over Euclidean metrics $g$
weighted by the Einstein-Hilbert action:
\begin{equation}
 S_E([g_{ab}]) \;= \frac{1}{16 \pi G} \;
 \int_{\cal M} \; {\rm d}^D \xi \;\sqrt{g(\xi)} (R - 2\Lambda),
\end{equation}
where $G$ is the Newton's gravitational constant, 
$\Lambda$ the cosmological constant and $R$ the scalar curvature
of the metric $g_{ab}$ on the space-time manifold ${\cal M}$.

Simplicial gravity, or dynamical triangulations, 
is a regularization of this model where the integration
over metrics is replaces by all possible gluings of $D$-simplexes 
into closed simplicial manifolds \cite{david,janbook}.  
In three dimensions
the regularized partition function becomes
\begin{equation}
Z(\mu,\kappa) \;=\; \sum_{T\in{\cal T}} \; \frac{1}{C_T} \;
{\rm e}^{\textstyle -\mu N_3 + \kappa N_0}
\label{model}
\end{equation}
where ${\cal T}$ denotes an appropriate ensemble of three-dimensional
simplicial manifolds, $C_T$ is the symmetry factor of a
triangulation $T$, $N_i$ is the number of
$i$-simplexes in the triangulation, 
and $\mu$ and $\kappa$ are coupling constants related 
to the cosmological and the inverse Newton's constant, 
respectively.

As it stands the model Eq.~(\ref{model}) needs not to be
well defined for any value of the couplings;
unless the sum over manifolds is restricted in some way
it is not convergent \cite{janbook}.
The same is true in two dimensions where, without the
restriction to triangulations of fixed topology, 
the number of triangulations
is not exponentially bounded as a function of volume --- 
a necessary condition for convergence of the sum.  
In three dimensions the situation is even worse,
there is no proof that even restricted to fixed topology 
the sum is convergent.  The existence of an
exponential bound is, 
however, strongly supported by numerical simulations
for an ensemble of spherical combinatorial triangulations 
\cite{jan3d,sim3d}. 

The model Eq.~(\ref{model}) has been studied extensively
using numerical simulations \cite{jan3d,sim3d,jap3d,jab3d2}. 
It has a discontinuous phase transition at a value of the 
inverse Newton's constant $\kappa \approx 3.8$, 
separating a strong-coupling (small $\kappa$)
crumpled phase from a weak-coupling branched polymer phase.
The crumpled phase is dominated by triangulations 
of almost zero extension; characterized by
few vertexes connected to a large portion of the
triangulation and has a internal fractal dimension $d_H = \infty$.
This excludes any sensible continuum limit in this
phase as the distance between two points will always stay 
at the Planck scale.  The branched polymer phase is, on the 
other hand, dominated by essentially ''one-dimensional`` 
triangulations  --- bubbles glued together along small necks 
into a tree-like structure  --- with fractal dimension two.   

Little attention has been paid so far to the ensemble of
triangulations included in the partition function Eq.~(\ref{model}),  
although in two dimensions it is well known that while different
ensembles yield the same continuum theory \cite{mat2} 
the finite-size effects depend strongly the
type of triangulations used.  
It has been shown that the less restricted the
triangulations are --- the larger
the ensemble of triangulations is --- the smaller the finite-size 
effects are \cite{medeg,minimal}.  
This result can be understood intuitively as, 
for a given volume, with a larger triangulation-space 
it is easier to approximate a particular fractal structure.  
This is especially important as in simulating models of 
dynamical triangulations
the geometric finite-size effects usually dominate. 

Simulations of simplicial gravity in dimensions larger than
two have until now used exclusively {\it combinatorial} triangulations.
In a combinatorial triangulation each $D$--simplex is uniquely 
defined by a set of ($D+1$) vertexes, i.e.\ it is combinatorially
unique.  In this paper I explore a different ensemble of 
triangulations, {\it degenerate} triangulations,
defined by relaxing this restriction and
allow {\it distinct} simplexes defined by the
same set of vertexes.  This allows, for example, two
vertexes connected by more than on edge and simplexes 
that have more than one face in common.  I retain, on
the other hand, the restriction that for every simplex 
all its ($D+1$) vertexes should
be different, i.e.\ I exclude {\it degenerate} simplexes.
As the ensemble of degenerate triangulation includes combinatorial
triangulations as a subclass, it is obviously larger and
as I will demonstrate in this paper this
leads to smaller finite-size effects, just as 
in two dimensions.  Apart from this, 
using degenerate rather than combinatorial triangulations
does not seem to change the phase structure of
the model Eq.~(\ref{model}); 
as the coupling constant $\kappa$ is varied I observer
a discontinuous phase transition separating a crumpled
phase from a branched polymer phase.

The paper is organized a follows:  In Section~2\; I define
the different ensembles of three-dimensional triangulations to
be considered and discuss some practical aspects 
such as ergodicity of the algorithms used in the
simulations and the possible problem with pseudo-manifolds.  
In Section~3\; I describe the numerical setup
and compare the efficiency of the simulations and 
the finite-size effects for the ensembles of 
degenerate {\it versus} combinatorial triangulations. 
In Section~4\; I investigate the phase structure of the
model, identify a phase transition and show that it is
discontinuous, and explore the different phases.
Finally, in Section~5\; I discuss the practical consequences
of this for further simulations of simplicial gravity.

\section{Degenerate triangulations}

\subsection{\sl Different ensembles of triangulations}

The triangulations included in the partition function
Eq.~(\ref{model}) are constructed by successively gluing 
equilateral 3-simplexes (tetrahedra) together 
along their two-dimensional faces (triangles) into a closed
three-dimensional simplicial manifold  \cite{david,janbook}.  
Pseudo-manifolds are eliminated from the ensemble
by the restriction that the neighborhood of every 
vertex in the triangulation should be homeomorphic to a sphere.  
To each tetrahedra there is associated a set of 
four vertexes, or 0-simplexes; the {\it order} of a vertex 
(or more generally of a sub-simplex) is defined as the
number of tetrahedra containing that vertex.  A {\it dual}
graph to a triangulation is defined by placing a vertex in
the center of each tetrahedra and connecting vertexes
in adjacent tetrahedra together.  For a three-dimensional
triangulation the dual is a $\phi^4$-graph.
A triangulation is said to be {\it degenerate} if it
contains two {\it distinct} 3-simplexes, or sub-simplexes, 
which are combinatorially identical, i.e.\ 
defined by the same set of vertexes.
A simplex is said to be degenerate if it is defined
by a set of vertexes including  the same vertex 
more than once.

By imposing additional restrictions on how the simplexes
may be glued together, it is possible to define different
ensembles of triangulations.  In this paper I will 
consider two such ensembles; combinatorial and degenerate
triangulations.  Before discussing these different
ensembles in three dimensions, let's start by describing the
corresponding triangulations in the simpler case
of two dimensions.  Given the definitions above
we distinguish among three types of 
two-dimensional triangulations:

\newpage

\begin{mydescription}

\item[(a)]
{\sl Combinatorial triangulations} : ${\cal T}_C$ \\ 
In a combinatorial triangulation every triangle, 
and every edge, is combinatorially distinct, i.e.\
it is uniquely defined by a set of distinct vertexes.  
From this it follows that no two triangles have more
than one edge in common and a triangle cannot be
its own neighbor.  The dual graph to a combinatorial
triangulation is a $\phi^3$-graph, excluding
both tadpoles and self-energy diagrams.

\item[(b)]
{\sl Restricted degenerate triangulations} : ${\cal T}_{D_R}$ \\
By allowing distinct triangles (or edges)
that are combinatorially equivalent, i.e.\ defined by the
same set of three (two) vertexes, while still excluding
degenerate triangles, we define a restricted ensemble
of degenerate triangulations, ${\cal T}_{D_R}$.
This type of triangulations can have vertexes
connected by more than one edge, and triangles that have
two edges in common. In this case self-energy diagrams
are allowed in the dual $\phi^3$-graph. 

\item[(c)]
{\sl Maximally degenerate triangulations} : ${\cal T}_{D_M}$ \\
Finally, by also allowing degenerate triangles we define
an ensemble of maximally degenerate triangulations, 
${\cal T}_{D_M}$.  This ensemble allows self-loops ---
edges starting and ending at the same vertex --- equivalently,
both tadpoles and self-energy diagrams are included
in the dual $\phi^3$-graph.

\end{mydescription}

\begin{figure}
\epsfxsize=4in \centerline{ \epsfbox{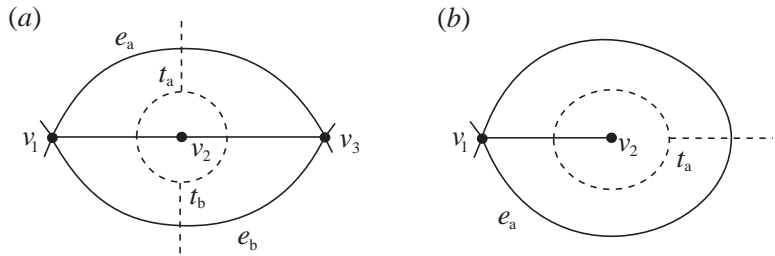}}
\caption[tri_2d]{\small An example of degenerate 
 two-dimensional triangulations.  The dashed lines indicate 
 the corresponding dual graphs.  
 In ({\it a}) the triangles $t_a$ and $t_b$ have two edges 
 in common, $\{v_1,v_2\}$ and $\{v_2,v_3\}$ 
 (a self-energy diagram);
 this type of degeneracy is allowed in restricted
 degenerate triangulations. In ({\it b}) the triangle
 $t_a = \{v_1,v_1,v_2\}$ is degenerate (corresponds to a tadpole); 
 this is allowed in maximally degenerate triangulations.}
\label{tri_2d}
\end{figure}

\noindent
Clearly 
${\cal T}_C \subset {\cal T}_{D_R} \subset{\cal T}_{D_M}$.
I show examples of those different types of degeneracy
in Figure~\ref{tri_2d}.   
As models of pure two-dimensional simplicial gravity 
defined with those different ensembles 
can been solved analytically as matrix models, it is known
that in all cases they define the same continuum theory \cite{mat2}.  
This is, of course, what one would expect based
on universality; the different ensembles 
simply amount to different discretization of the manifolds
and the details of the discretization should be 
irrelevant in the continuum limit.  This is true
even if the triangulations are restricted even further and 
only minimal curvature fluctuations of $\pm 1$
are allowed (including only vertexes of order 5, 6 and 7) 
\cite{minimal}.  But, as mentioned in the Introduction,
the finite-size effects do depend on the ensemble
used, in particular they increase as more restrictions
are placed on the triangulations \cite{medeg,minimal}.

\begin{figure}
\epsfxsize=4in \centerline{ \epsfbox{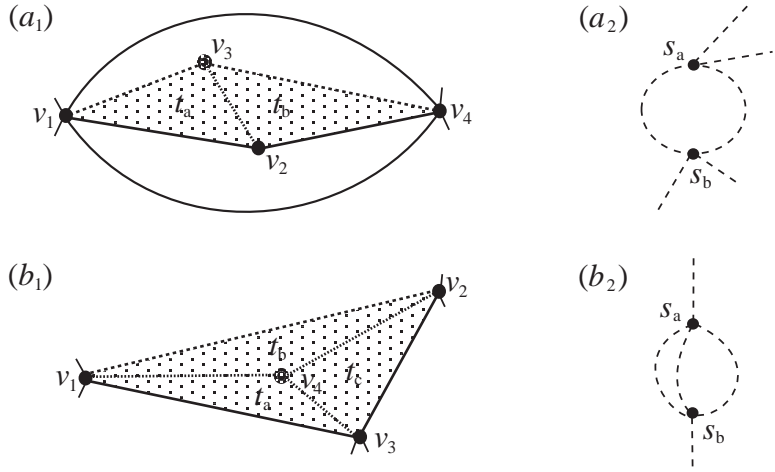}}
\caption[tri_3dA]{\small Two examples of degeneracy 
 allowed in three-dimensional degenerate triangulations. 
 Figures ($a_2$) and ($b_2$)
 show the corresponding dual $\phi^4$-graphs.  
 In both examples there are two distinct but combinatorially
 equivalent tetrahedra $s_a$ and $s_b$, 
 defined by the same set of vertexes $\{v_1,v_2,v_3,v_4\}$.  
 In ($a_1$) the tetrahedra have two triangles in common, whereas
 in ($b_1$) they share three triangles, in which case 
 the dual graph is a self-energy diagram.}
\label{tri_3dA}
\end{figure}

Analogous to two dimensions, by imposing
the appropriate restrictions in three dimensions
I defined the corresponding three different ensembles
of triangulations. By allowing distinct but 
combinatorially equivalent non-degenerate simplexes,
I define an ensemble of restricted degenerate
three-dimensional triangulations.  Compared to
combinatorial  triangulations, in this ensemble
two tetrahedra may be glued together along more than
one of their faces.  I show examples of this 
in Figure~\ref{tri_3dA}.

Relaxing these 
restrictions further and allow degeneracy within
the simplexes themselves defines an  
ensemble of maximally degenerate triangulations\footnote{Note that
in contrast to two (or more generally even) dimensions
it is not possible to construct a three-dimensional triangulation
with tadpoles in the corresponding dual $\phi^4$-graph.  For a part of
the dual graph to be connected to the rest by a single link,
the corresponding part of the triangulation would have to
be enclosed by a (closed) boundary consisting of a single
triangle.  This is, however, impossible.}, ${\cal T}_{D_M}$
This allows, for example, self-loops 
both in the triangulation and in its dual graph
(a simplex that is its own neighbor).
Although this type of degenerate triangulations is {\it a priori}
well defined, my numerical investigation indicates
that this ensemble does not lead to a well 
defined statistical model.  
I tried to simulate the model Eq.~(\ref{model})
using maximally degenerate triangulations;
however, in quasi-canonical simulations (see Section~3)
it was not possible to tune the cosmological constant 
$\mu$ in such a way that the volume fluctuated around a 
fixed target volume $\bar{N}_3$.  
Regardless of the value of $\mu$ used, in the course of the 
simulations the volume eventually exploded.
This behavior suggests that the numbers of maximally 
degenerate three-dimensional triangulations grows faster than 
exponentially with the volume, in which case the partition 
function Eq.~(\ref{model}) does not converge.
In the rest of this paper I only consider the ensemble 
${\cal T}_{D_R}$, omitting the specification restricted
when referring to it.

\begin{figure}
\epsfxsize=4in \centerline{ \epsfbox{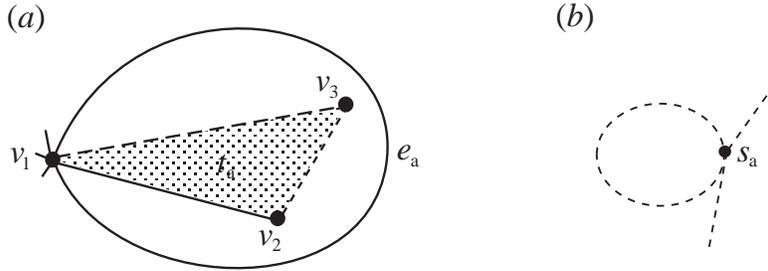}}
\caption[tri_3dB]{\small An example of a degenerate simplex,
 $s_a = \{v_1,v_1,v_2,v_3\}$, allowed in a three dimensional
 maximally degenerate triangulation.  The edge $e_a = \{v_1,v_1\}$
 is a self-loop.}
\label{tri_3dB}
\end{figure}

\subsection{\sl Ergodicity in the updating procedure}

Having defined degenerate triangulations,
a necessary prerequisite for using them in numerical
simulations of simplicial gravity is that
there exist a set of ergodic geometric moves which
allow us to explore the triangulation-space.
For combinatorial triangulations this is provide
by the ($p,q$)--moves, a variant of the 
Alexander moves \cite{move}. 
In three dimensions the ($p,q$)--moves consist of 
either inserting or removing a vertex of order four, or
replacing a  triangle by an edge of order three and vice verse. 
These moves are depicted in Figure~\ref{movefig}.
To apply the same set of moves in simulations with degenerate
triangulations I have to show that they are
also ergodic when applied to the ensemble  ${\cal T}_D$.
To do so it is sufficient to show that every degenerate 
triangulation can be changed into a combinatorial triangulation 
using a finite sequence of the ($p,q$)--moves.  
I distinguish among three different types of 
degeneracy in restricted degenerate triangulations: 
combinatorially equivalent {\it tetrahedra}, 
{\it triangles} and {\it edges}, all of which can be
eliminate following the appropriate procedures:

\begin{figure}
\epsfxsize=5.5in \centerline{ \epsfbox{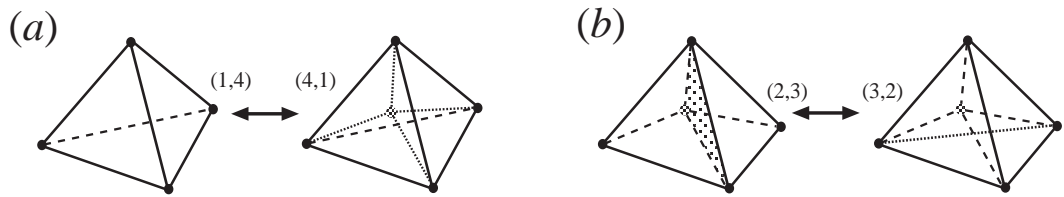}}
\caption[movefig]{\small The ($p,q$)--moves in three dimensions:
 ({\it a}) Inserting a vertex and its inverse, deleting a vertex. 
 ({\it b}) Replacing a triangle by an edge  and vice verse.}
\label{movefig}
\end{figure}

\begin{mydescription}

\item[(a)]
{\sl  Equivalent tetrahedra}: \\
In the case of two, or more, tetrahedra defined by the same
set of vertexes it is sufficient to insert a vertex
(applying move $(1,4)$) into each of those tetrahedra,
this will create a set of combinatorially distinct 3-simplexes.

\item[(b)]
{\sl Equivalent triangles}: \\
Given two triangles defined by identical set of vertexes I
can apply move $(2,3)$ to one of them, i.e.\ 
replace that triangle with a new edge. To ensure that the
new edge does not result in multiply connected vertexes, and
to make sure the move is not rejected by a geometric
constraint, prior to using move $(2,3)$ I can apply 
move $(1,4)$ to both of the tetrahedra containing the 
triangle in question --- the new edge now joins two 
previously unconnected vertexes.

\item[(c)]
{\sl Equivalent edges}: \\
It is slightly more complicated to eliminate a par
of distinct edges connecting the same two vertexes.
To remove an edge I must apply move $(2,3)$, i.e.\
replace it with a triangle.  This is, however, only possible
if the corresponding edge is of order three.
To reduce the order of the edge 
I apply move $(2,3)$ repeatedly to the neighboring triangles.
It is possible that some of those triangles 
could be combinatorially identical; this I eliminate
by first applying procedure ({\sf b}).  The remaining
triangles can then safely be replaced until the order of
the edge is three. 
 
\end{mydescription}

An additional complication that could arise when restrictions
on the triangulations is relaxed is the appearance of 
{\it pseudo-manifolds} \cite{david}.  In general when
gluing $D$--simplexes together into a closed manifold
one does not get a simplicial manifold but a pseudo-manifold.
A simplicial manifold is defined by the additional requirement 
that the neighborhood of each vertex is homeomorphic to the
$D$--dimensional ball $B_D$.
As this requirement is trivially satisfied for $D = 2$, 
all degenerate triangulations in two dimensions are 
automatically simplicial manifolds.
It is {\it a priori} not clear that in simulating
degenerate triangulations in three dimensions, pseudo-manifolds
could not be generated --- something we would like to avoid.

I observe, however, that in order
to generate a pseudo-manifold the ($p,q$)--moves must
change the topology of the dual complex to a 
given vertex $v_0$ in the triangulation.
By the dual complex I mean the two-dimensional
triangulation constructed by connecting by edges 
the centers of all adjacent tetrahedra containing $v_0$.
But as the moves only change the triangulation,
hence the dual complex, locally they will not change its
topology.  More precisely, when applying a ($p,q$)--move 
to a three-dimensional triangulation, degenerate or not,
the change in the local neighborhood of each of the vertexes
involved in the move can be given in terms of
the corresponding two-dimensional ($p,q$)--moves 
(i.e.\ inserting/deleting a vertex or flipping an edge) 
applied to its dual complex.  This ensures that the
topology of the dual complex is not changed when
the move is applied and pseudo-manifolds are not
generated.

\section{Simulating degenerate triangulations}

To study the model Eq.~(\ref{model}) I now turn to
numerical simulations.    
As the ($p,q$)--moves in three dimensions do not
preserve the volume of the triangulation 
the simulations cannot be restricted to the canonical ensemble, 
contrary to what is possible  in two dimensions. 
In addition, fluctuations in the
volume are necessary to maintain ergodicity in 
the updating procedure.  
In practice though I perform quasi-canonical Monte Carlo 
simulations, with almost fixed $N_3$, simulating the
partition function
\begin{eqnarray}
 Z(\mu,\kappa;\bar{N}_3) &\;=\;
 &\sum_{N_3} \; {\rm e}^{ \textstyle - \mu N_3 - 
 \gamma (N_3 - \bar{N}_3)^2} \;
 \Omega_{N_3}(\kappa), \\
 \Omega_{N_3}(\kappa) &\;=\; 
 &\sum_{T\in{\cal T}({N_3})} \;\frac{1}{C_T} \;
 {\rm e}^{\textstyle \; \kappa N_0}.
 \label{quasi}
\end{eqnarray}
Here $\Omega_{N_3}(\kappa)$ is the canonical partition
function and the quadratic potential term added to
the action ensures, for an appropriate choice of
the parameter $\gamma$, the necessary fluctuations around
a target volume $\bar{N}_3$.  
In simulating degenerate triangulations I found
$\gamma \approx 0.001$ to be an appropriate choice.  
In all cases I simulate an ensemble of spherical
triangulations.

From a practical point of view simulating
degenerate triangulations is actually simpler than simulating
combinatorial triangulations.  For the latter the most
time-consuming part of the update is the manifold
check, i.e.\ to verify that there exist no sub-simplex
in the triangulation combinatorially equivalent to the one 
created by the move \cite{simprog}.  
In three dimensions this check is done both when a triangle
is replaced by an edge, and vice verse.
This check is not needed for degenerate triangulation; the
only geometric check that remains is in replacing
a triangle by an edge the two tetrahedra involved 
must only have one triangle in common.
As this check is local the computational effort needed
to update a degenerate triangulation is less than
for a combinatorial one.

This does not automatically imply that 
simulations with degenerate triangulations are more efficient.  
As the ensemble of degenerate triangulations
is much larger the corresponding critical cosmological constant 
$\mu_c$ is also larger than for the ensemble of combinatorial 
triangulations. And as $\mu_c$ enters the detail balance condition
used in the Metropolis test 
this in effect reduces the acceptance rate in 
the updating procedure.
To compare the efficiency of simulating the
different ensembles I must estimate how fast, measured
in CPU-time, independent configurations are generated.  
I have done this for both ensembles
in the crumpled phase, $\kappa = 0$.
I caution, however, that the comparative efficiency
will depend on several factors: the particular implementation of
the algorithm, how efficiently the manifold checks are 
executed, the probability of choosing the different moves,
and where in the phase-space the simulations are carried
out.  In the comparison I use the same program for
both types of triangulations, turning on and off the 
manifold checks, and choose the different moves with
equal probability.  I measure the CPU-times $t_s$ (in ms)
it takes to perform one ''sweep`` through the
triangulations; a sweep is defined as $\bar{N}_3$ attempted
local moves, and the auto-correlation times $\tau$ in units 
of sweeps, measured for the time-series of the 
energy-density $n_0 = N_0/N_3$.  Combining $t_s$ and $\tau$, 
I get an estimate of how fast independent configurations are 
generated: 
\begin{equation}
 {\cal T}_{n_0} \;\approx\; \left \{ \begin{array}{ll} 
   0.43 \; N_3^{\; 2.07} & \mbox{combinatorial,} 
   \vspace{5pt} \\ 
   230 \;\; N_3^{\; 1.14} & \mbox{degenerate.}  
 \end{array} \right.
 \label{perf1}
\end{equation}
For small triangulations, $N_3 \approx 800$, the effective 
auto-correlations are comparable for the two ensembles,
whereas for larger volumes simulating
degenerate triangulations is substantially faster.   

\begin{figure}
\epsfxsize=4in \centerline{ \epsfbox{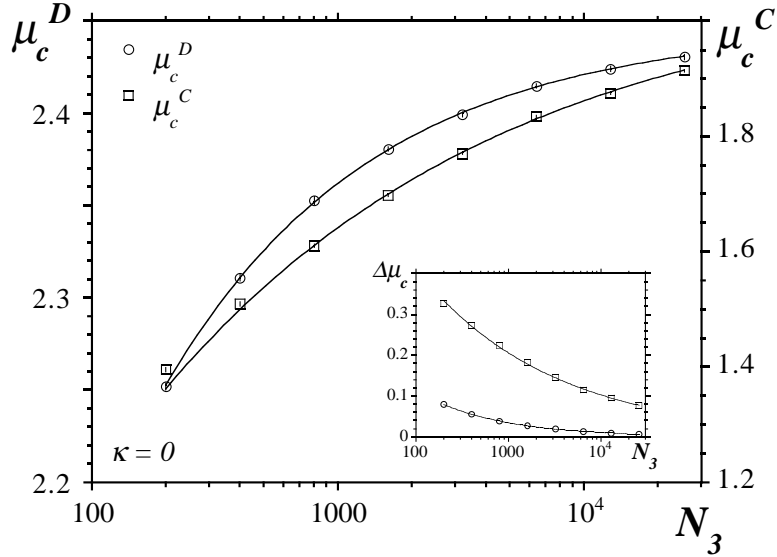}}
\caption[cosmo]{\small The effective critical cosmological
 constant $\mu_c(N_3)$, both for an ensemble of 
 degenerate (circles) and combinatorial (diamonds) 
 triangulations.  The solid lines are fit to the
 form: $\mu_c(N_3) = \bar{\mu}_c + \beta/N_3^{\gamma}$.
 {\sf Insert}: The relative deviation from the
 infinite volume value: $\Delta \mu_c = |\bar{\mu}_c
 - \mu_c(N_3) |/\bar{\mu}_c$.}
\label{cosmo}
\end{figure}

The real benefit of using degenerate triangulations
comes from the reduction in finite-size effects.  This
reduction is especially pronounced in the crumpled
phase where the internal
geometry is dominated by collapsed manifolds.
To demonstrate this I measured the volume dependence
of the effective critical cosmological constant $\mu_c^D(N_3)$
for $\kappa = 0$ and for triangulations of volume
$N_3 \lesssim 25000$.
This I show in Figure~\ref{cosmo}.  For comparison I
include in the figure the corresponding values
$\mu_c^C(N_3)$ for combinatorial triangulations.
For both ensembles $\mu_c(N_3)$ approaches a
finite value $\bar{\mu}_c$ as the volume diverges; 
this in turn implies that the canonical partition function 
$\Omega_{N_3}$ is exponentially bounded as a function of
volume --- an necessary
condition for the partition function Eq.~(\ref{model}) to
be convergent.  I have fitted the data to an assumption of
a power-law convergence to an exponential bound 
(including volumes $N_3 \geq 400$); this yields:
\begin{equation}
\mu_c(N_3) \;=\; \left \{ \begin{array}{llll} 
 2.073(21) - 3.67(44)\; N_3^{\; -0.31(2)}, 
 & \chi^2 = 4.0 \qquad \mbox{combinatorial,} 
 \vspace{5pt} \\ 
 2.447(3) \;\;- 2.97(29)\;N_3^{\;-0.51(2)}, 
 & \chi^2 = 0.8 \qquad \mbox{degenerate.}  
 \end{array} \right.
 \label{expo}
\end{equation}
For both ensembles the quality of the fit (the $\chi^2/{\rm d.o.f}$)
is acceptable.  This can be compared to a fit to 
a super-exponential growth of $\Omega_{N_3}$,
compatible with the assumption $\mu_c(N_3) \approx \alpha 
+ \beta \log N_3$, which is definitely ruled out by a very
large $\chi^2$ value.

There is, however, a marked difference in how
fast $\mu_c$ approaches its infinite
volume value for those two ensembles.
This is shown in the insert in Figure~\ref{cosmo} where I
plot $\Delta \mu_c = |\bar{\mu}_c - \mu_c(N_3) |/\bar{\mu}_c$.  
Taken at face value, this suggests 
a reduction of finite-size effects by two orders of 
magnitude when using degenerate rather than combinatorial
triangulations.   This huge reduction is though presumable
only achieved in the crumpled phase where collapsed manifolds
dominate the partition function and the
geometric finite-size effects are most pronounced.

\section{The phase structure}

\subsection{\sl Existence of a phase transition}

I now turn to the existence of a phase transition
in the model as the inverse Newton's constant $\kappa$ is varied.
For combinatorial triangulations it is well established
by numerical simulations that the model Eq.~(\ref{model})
has a discontinuous phase transition at $\kappa \approx 3.8$,
separating a strong-coupling (small $\kappa$) crumpled phase from 
a weak-coupling branched polymer phase \cite{jan3d,sim3d,jap3d}.
As I will demonstrate in this section this is
also the case for degenerate triangulations; moreover,
the discontinuous nature of the transition is easily
observed on triangulations of relatively modest size.

\begin{figure}
\epsfxsize=4in \centerline{ \epsfbox{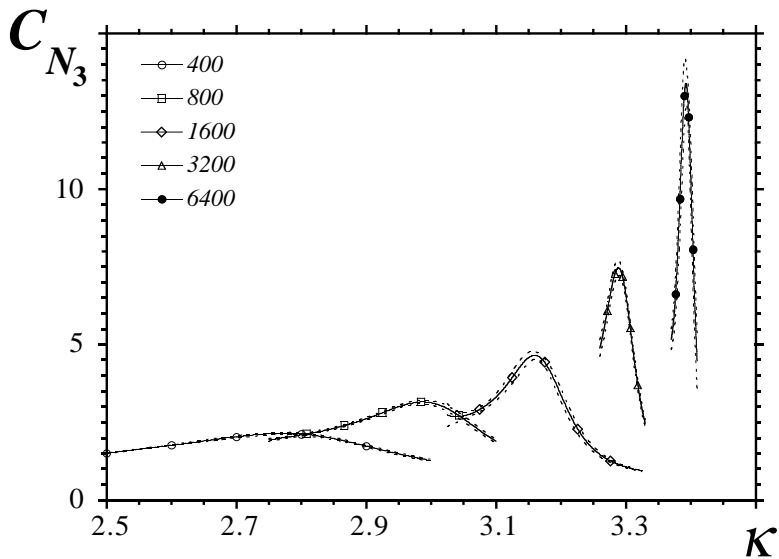}}
\caption[cv_n0]{\small The measured specific heat $C_{N_3}$ 
 --- the fluctuations in the number of vertexes $N_0$ --- 
 as the inverse Newton's constant $\kappa$ is varied.  
 This is for target volume $N_3 = 400$ to 6400.
 The solid lines are interpolations between measurements, 
 using multi-histogram methods, and the dashed lines indicate 
 the error.}
\label{cv_n0}
\end{figure}

I have simulated the model Eq.~(\ref{quasi}) for target
volumes $N_3 = 200$ to 6400, and for each volume 
I search for a phase transition in the fluctuations
of the energy-density $n_0 = N_0/N_3$.  A signal for a transition 
would be a peak in the specific heat:   
\begin{equation}
C_{N_3} \;=\; \frac{\kappa^2}{N_3} \;
 \left ( \langle N_0^2 \rangle  - \langle N_0 \rangle^2 \right ).
\end{equation}
This is shown in Figure~\ref{cv_n0} for different 
target volumes.  I did simulations at few values
of $\kappa$ around the observed peaks, collecting between
two and ten thousand independent measurements,
and then used standard multi-histogram methods to
interpolate between the measurements.  The interpolations
is shown as solid lines in Figure~\ref{cv_n0}, the dashed lines 
indicate the errors.

\begin{figure}
\epsfxsize=4in \centerline{ \epsfbox{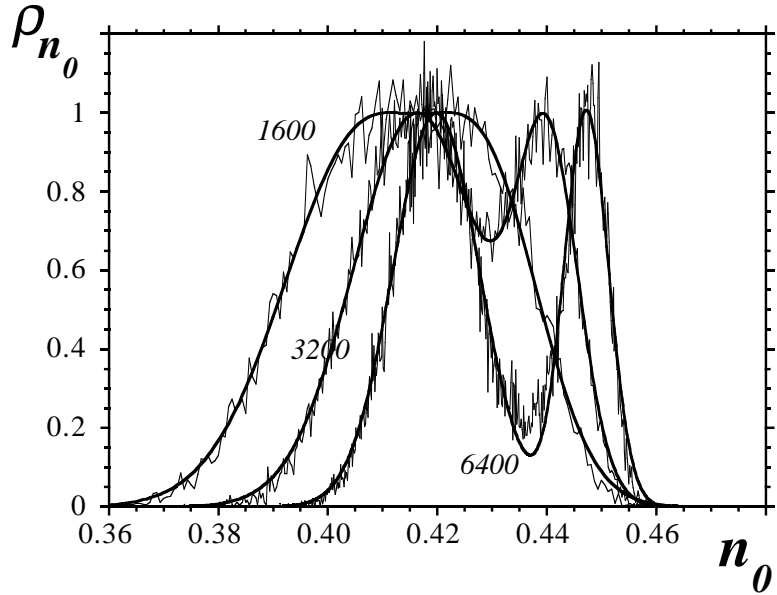}}
\caption[rho_n0]{\small The energy-density distribution
 $\rho(n_0)$ for $N_3 = 1600$, 3200 and 64000.  The thin
 lines are the measured distributions, the
 thick lines are fits to two Gaussian peaks Eq.~(\ref{twogaus}).  
 Each distributions is normalized with the height of the
 peaks and the values of $\kappa$ chosen so the two peaks 
 are of equal height.}
\label{rho_n0}
\end{figure}

To infer about the nature of the transition 
I look at the energy-density distribution 
$\rho(n_0)$. For sufficiently large target 
volume this distribution has a well resolved double-peak 
structure, indicating a discontinuous phase transition.
This double-peak structure is observed already on  
volumes $N_3 > 1600$.  I show an example of the
measured energy distributions in Figure~\ref{rho_n0} (thin lines) 
for the three largest volumes I studied, together with fits to a 
form composed of two Gaussian peaks (thick lines): 
\begin{equation}
 \rho(n_0) \;=\; a_1 \; {\rm e}^{\textstyle -c_1(b_1-n_0)^2}
   +a_2 \; {\rm e}^{\textstyle -c_2(b_2-n_0)^2}.
 \label{twogaus}
\end{equation}
Each distribution is normalized by the height of the peaks
and the value of $\kappa$ is chosen so that the two peaks are 
of equal height.

\begin{figure}
\epsfxsize=4in \centerline{ \epsfbox{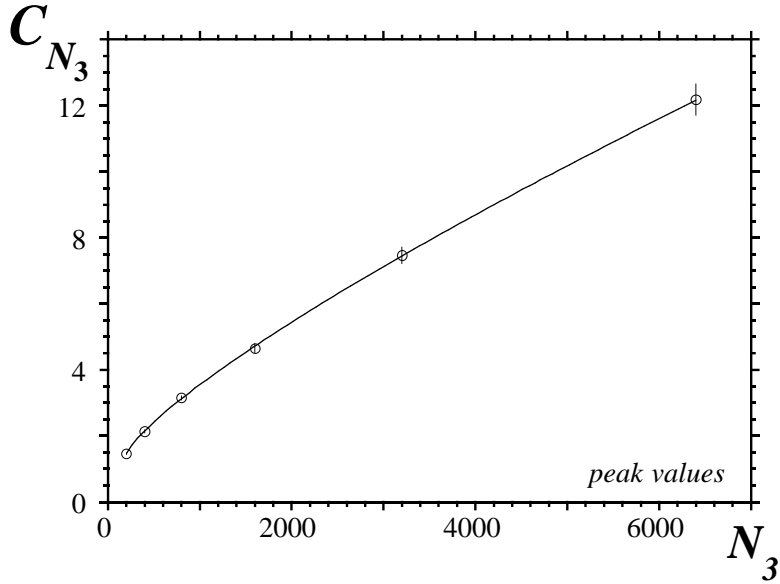}}
\caption[cv_peak]{\small The volume scaling of the peak in
 the specific heat.  The solid line is a fit to the scaling
 form: $C_{N_3} \approx a + b N_3^{\alpha}$.  The value of
 the exponent $\alpha$ is shown in Table~1.}
\label{cv_peak}
\end{figure}

Additional evidence for a phase transition comes from other    
geometric observables such as the maximal order of a vertex, 
$q_0$.  As discussed in the 
Introduction, the crumpled phase is dominated by
triangulations containing few vertexes of very large order,
whereas in a branched polymer phase the vertex orders are
more or less equally distributed.
This is reflected in a sudden drop in
the maximal vertex order across the transition; this change
leads to divergent peaks in both the fluctuations 
of $q_0$, i.e.\ the susceptibility $\chi_{q_0} = N_3 \;(\langle
q_0^2 \rangle - \langle q_0 \rangle^2)$, and its energy
derivative  ${\rm D}_{q_0} = N_3 \; (\langle q_0 n_0 \rangle
- \langle q_0 \rangle \langle n_0 \rangle)$.
In fact, plots of both $\chi_{q_0}$ and ${\rm D}_{q_0}$ look more or
less identical to Figure~\ref{cv_n0}.  For all the observables
the peak heights diverge, an example of this is shown
in Figure~\ref{cv_peak} for the specific heat.  I have 
estimated the corresponding scaling behavior, fitting the
peak values to the form: ${\cal O}_{peak} \sim N_3^{\alpha}$;
the fitted scaling exponents are shown in Table~1.  
Compared to $\chi_{q_0}$ and ${\rm D}_{q_0}$, the
fit of the specific heat is rendered considerable
more difficult by an additional constant contribution to
the scaling behavior.  Nevertheless, for the specific
heat the peak height scales almost linearly for sufficiently
large target volume
as is expected for a discontinuous phase transition.

\begin{figure}
\epsfxsize=4in \centerline{ \epsfbox{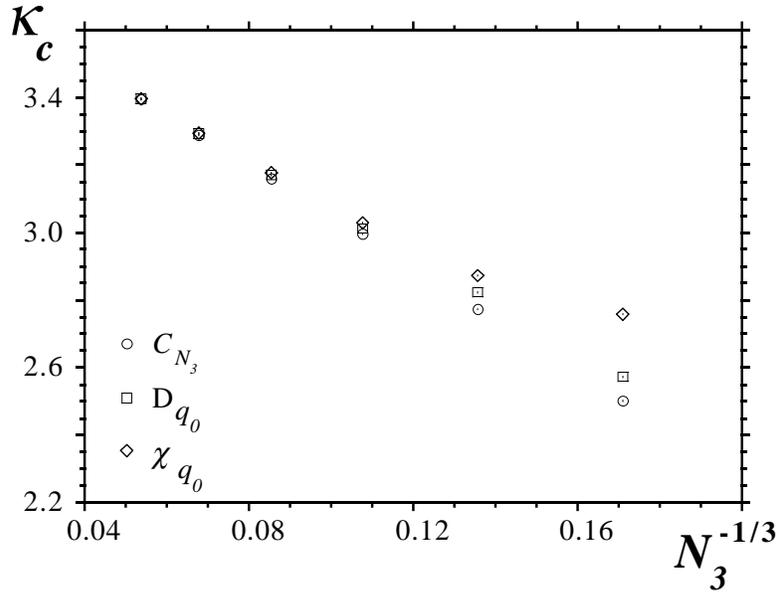}}
\caption[kap_c]{\small Scaling of the pseudo-critical
 inverse Newton's constant $\kappa_c(N_3)$, defined 
 by the location of peaks in $C_{N_3}$, $\chi_{q_0}$ and 
 ${\rm D}_{q_0}$ respectively.  
 The values are plotted {\it versus}
 the volume $N_3$ scaled with an exponent $x=1/3$.}
\label{kap_c}
\end{figure}

To locate the infinite volume critical coupling $\bar{\kappa}_c$
I fit the pseudo-critical couplings $\kappa_c(N_3)$, defined
by the location of peaks in the different observables, 
to the expected finite-volume scaling behavior:
\begin{equation}
\kappa_c(N_3) \;=\; \bar{\kappa}_c + \frac{const.}{N_3^x}.
 \label{coupscal}
\end{equation}
The fitted values of $\bar{\kappa}_c$ and the exponent $x$
are shown in Table~1.  To asses the finite-size effects I 
repeated the fits with a different lower cut-off on the
included target volume. This did not affect the estimate
of $\bar{\kappa}_c$ appreciatively; in all
cases  $\bar{\kappa}_c \approx 3.8$. Note however that,
in contrast to what has recently been reported in simulations 
with combinatorial triangulations \cite{jap3d}, I
observe no indication of the peak locations running
to $\bar{\kappa}_c = \infty$.  
For all the observables the pseudo-critical
couplings converge nicely.
This is shown in Figure~\ref{kap_c}.
A fit to an assumption of non-convergence, 
$\kappa_c(N_3) = a + b \log N_3$, is
ruled out by a $\chi^2/({\rm d.o.f.}) \gtrsim 1000$.
For the scaling exponent $x$, I get values $x \approx 1/3$.
For a discontinuous phase transition one would 
naively expect 
$\delta\kappa_c = |\kappa_c(N_3) - \bar{\kappa_c}| 
 \sim L^{-d_H}$, where $L$ is some appropriate length
scale in the system and $d_H$ is the (internal) fractal
dimension (see e.g.\ Ref.~\cite{binder}).  
This would imply $x = 1$ in Eq.~(\ref{coupscal}).
However, for a system with a fluctuating
internal geometry like simplicial gravity, it is not
clear what dimensionality should be associated to the
system at a discontinuous phase transition where in
one phase the internal fractal dimension is infinite,
but two in the other.  Hence it is not obvious
how to interpret the observed value of 
the exponent $x$.  

\begin{table}
 \begin{center}
 \label{tabbc}
 \caption{\small ({\sf A}) The volume scaling of the
  peaks in $C_{N_3}$, $\chi_{q_0}$ and ${\rm D}_{q_0}$, fitted
  to the form: $a + b \bar{N}_3^{\alpha}$, together with
  the quality of the fits.  For
  $\chi_{q_0}$ and ${\rm D}_{q_0}$ I use $a=0$.  
  Values are shown for different lower cut-off $N_3^{min}$ 
  on the target volume include in the fit.
  ({\sf B}) The corresponding values of $\bar{\kappa}_c$, 
  determined from fitting the pseudo-critical coupling constants 
  $\kappa_c(N_3)$ to the form Eq.~(\ref{coupscal}),
  together with the scaling exponent $x$.}
 \vspace{8pt}
 {\small
 \begin{tabular}{|l|llc|llc|llc|} \hline
 \vspace{-5pt} & & & & & & & & &    \\
  $N_3^{min}$
   & \multicolumn{3}{|c|}{$C_{N_3}$}
   & \multicolumn{3}{c|}{$\chi_{q_0}$}
   & \multicolumn{3}{c|}{${\rm D}_{q_0}$} \\   
 & & & & & & & & &   \\  
 {\large({\tt A})}  & & & & & & & & & \\  
   & \multicolumn{2}{|c}{$\alpha$} & $\chi^2$ 
   & \multicolumn{2}{|c}{$\alpha$}  & $\chi^2$
   & \multicolumn{2}{|c}{$\alpha$}  & $\chi^2$ \\ \hline
 \vspace{-5pt} & & & & & & & & &   \\ 
 $  200$  &  \multicolumn{2}{|c}{0.80(5)} &  2.5
          &  \multicolumn{2}{|c}{0.78(2)} &  3.7
          &  \multicolumn{2}{|c}{0.66(4)} &  0.7 \\	  
 $  400$  &  \multicolumn{2}{|c}{0.80(7)} &  1.0 
          &  \multicolumn{2}{|c}{0.77(2)} &  2.8 
          &  \multicolumn{2}{|c}{0.66(5)} &  0.5 \\	  
 $  800$  &  \multicolumn{2}{|c}{0.85(12)}&  0.5 
          &  \multicolumn{2}{|c}{0.76(1)} &  0.6 
          &  \multicolumn{2}{|c}{0.66(7)}&  0.3 \\ 	  
  & & & & & & & & &    \\ 
 {\large ({\tt B})} & & & & & & & & & \\  
   & $\bar{\kappa}_c$ & $x$ & $\chi^2$
   & $\bar{\kappa}_c$ & $x$ & $\chi^2$
   & $\bar{\kappa}_c$ & $x$ & $\chi^2$ \\ \hline
 \vspace{-5pt} & & & & & & & & &   \\ 
 $  200$  &  3.755(22) &  0.36(2) &  8.5 
          &  3.790(19) &  0.32(2) &  4.8
          &  3.755(22) &  0.27(5) &  24  \\	  
 $  400$  &  3.741(37) &  0.37(3) &  7.5 
          &  3.829(22) &  0.31(2) &  2.3
          &  4.10(14)  &  0.22(5) &  12  \\	  
 $  800$  &  3.95(18) &  0.27(7) &  8.5 
          &  3.49(27) &  0.26(7) &  2
          &  4.07(43) &  0.21(9) &  11   \\ 	  
\vspace{-5pt} & & & & & & & & &  \\	  \hline
 \end{tabular} }
 \end{center}
\end{table}

\subsection{\sl The crumpled phase}

As already stated the strong-coupling phase of the
model Eq.~(\ref{model}) is a crumpled phase, characterized
by few vertexes of very large order.  Qualitatively
I observe this behavior regardless
of the type of triangulations I simulated.  
Note, however, that there is {\it a priori} no reason to expect
exactly the same behavior for the two ensembles in 
this phase; because of the collapsed nature of manifolds
it is unlikely that any sensible continuum limit 
can be taken for $\kappa < \kappa_c$.  Hence
the details of the triangulations --- the discretization ---
may be relevant in the infinite volume limit.

There does indeed appear to be some slight difference in the
internal geometry of the ensemble of degenerate triangulations,
compared to the combinatorial ensemble, for $\kappa = 0$.  
An example of this is the scaling behavior of the 
effective critical cosmological constant determined in Section~3.
For combinatorial triangulations the exponent of the
power-law correction is $\gamma \approx 0.3$, whereas
$\gamma \approx 0.5$ for degenerate triangulations.
Similar difference is observed in the volume scaling
of the order of the most singular vertex. For
three-dimensional combinatorial triangulations $q_0$
scales sub-linearly with volume: $q_0 \sim N_3^{0.42(2)}$,  
for degenerate triangulations, on the other hand, the scaling
exponent is larger; $q_0 \sim N_3^{0.60(5)}$.  

The collapsed nature of this phase is most easily 
demonstrated by measuring the fractal dimensions $d_H$.
To estimate $d_H$
I measure the two-point correlation function $g_{N_3}(r)$,
defined as the number of simplexes at a geodesic distance
$r$ from a marked simplex.  The geodesic distance between
two simplexes is defined as the shortest path
traversed through the center of simplexes.
Assuming that the only relevant length scale in the system is 
given by $N_3^{1/d_H}$, general
scaling arguments \cite{scaling} imply that
\begin{equation}
 g_{N_3}(r) \;\sim\; N_3^{1-1/d_H} \; F(x), \qquad
 x \;=\; \frac{r}{N_3^{1/d_H}}.
 \label{scale}
\end{equation}
The fractal dimension is measured by ''collapsing`` 
distributions $g_{N_3}(r)$, corresponding to different volumes, 
onto a single scaling curve.
I have done this for volume $N_3 \lesssim 6400$ and for $\kappa = 0$;
the value of $d_H$, determined by minimizing the $\chi^2$--value
of the collapse, is shown in Table~2.
The errors are estimated using jack-knifing.
To demonstrate how the estimate of $d_H$ depends on the volume,
I have done the analysis either including all the volumes,
or using pairs of consecutive volumes $N_3$ and $2N_3$.
For $\kappa = 0$ the estimate of $d_H$
drifts systematically toward larger values as
the volume is increased, in addition the fit to the scaling
behavior Eq.~(\ref{scale}) is very poor; the $\chi^2$/(d.o.f.) $>> 1$.  
This is not unexpected,
in a crumpled phase the internal fractal dimension
is $d_H = \infty$, in which case the simplex-simplex correlation
function is not expected to scale.

\begin{table}
 \begin{center}
 \label{dh}
 \caption[dh]{\small The fractal dimension $d_H$ for
 an ensemble of degenerate three-dimensional triangulations,
 determined by collapsing the simplex-simplex correlation function
 $g_{N_3}(r)$ onto the scaling form Eq.~(\ref{scale}).  
 The values shown are for $\kappa = 0$ in the
 crumpled phase, and $\kappa = 5$ and 8 in the branched polymer
 phase.  The analysis is done including different target
 volumes $N_3$.  Also included is the
 $\chi^2/({\rm d.o.f.})$ value of the collapse.}
 \vspace{8pt}
 {\small
 \begin{tabular}{|c|cr|cr|cr|} \hline
 \vspace{-5pt} & & & & & &    \\
 &  \multicolumn{2}{|c|}{$\kappa = 0$} &
    \multicolumn{2}{c|}{$\kappa = 5$} &
    \multicolumn{2}{c|}{$\kappa = 8$} \\
  \{$N_3^{min}$---$N_3^{max}$\}
    & $d_H$ & $\chi^2$ & $d_H$ &  $\chi^2$
    & $d_H$ &  $\chi^2$  \\
 \vspace{-5pt} & & & & & &    \\  \hline
 \vspace{-5pt} & & & & & &    \\ 
  \{200 --- 400\}    & 4.05(15)  & 55
                     & 1.85(6)   & 0.64
                     & 1.89(5)   & 0.55 \\
  \{400 --- 800\}    & 4.66(18)  & 62
                     & 1.92(8)   & 0.84
                     & 1.95(7)   & 0.55 \\
  \{800 --- 1600\}   & 5.27(14)  & 50 
                     & 1.93(8)   & 0.77  
                     & 1.93(13)  & 0.32 \\
  \{1600 --- 3200\}  & 5.68(16)  & 32
                     & 2.00(13)  & 0.45  
                     & 1.98(15)  & 0.58 \\
  \{3200 --- 6400\}  & 6.25(20)  & 30 
                     & 1.99(15)  & 0.45
		     & 1.96(12)  & 0.50 \\
 \vspace{-5pt} & & & & & &   \\ 
  \{400 --- 6400\}   & 5.20(30)  & 84  
                     & 1.94(10)  & 1.1 
		     & 1.95(11)  & 0.6  \\  	  
 \vspace{-5pt} & & & &   & &  \\	  \hline
 \end{tabular} }
 \end{center}
\end{table}

\subsection{\sl The branched polymer phase}

\begin{figure}
\epsfxsize=4in \centerline{ \epsfbox{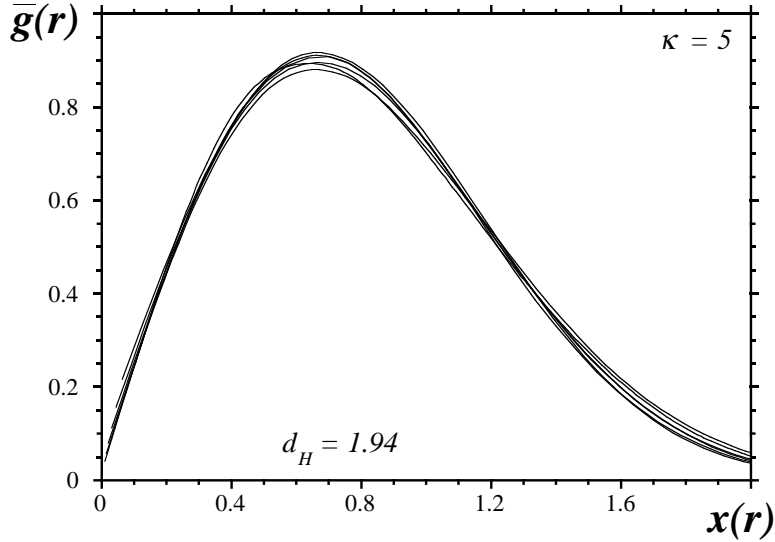}}
\caption[dhfig]{\small The scaled simplex-simplex correlation
 function $\bar{g}_{N_3}(x) = g_{N_3}(x)/N_3^{1-1/d_H}$
 {\it versus} the scaling variable $x = r/N^{1/d_H}$.  This is for
 degenerate three-dimensional triangulations of size
 $N_3 = 400$ to 6400 and $\kappa = 5$.  
 The measured value of $d_H = 1.94$ has been used in the
 scaling.}
\label{dhfig}
\end{figure}

Finally I have also investigated the nature of the
weak-coupling phase for degenerate triangulations at
two values of the coupling constant: $\kappa = 5$
and 8.  To establish that this is a branched polymer phase
I measured both the fractal dimension $d_H$ and the string
susceptibility exponent $\gamma_s$.
The estimate of $d_H$ is included Table~2. 
In contrast to the crumpled phase in this phase I get
a consistent value $d_H \approx 2$, 
as expected for branched polymers; 
for both values of $\kappa$ the quality of the scaling
is very good, $\chi^2/{\rm (d.o.f.)} \lesssim 1$.
An example of the scaling is shown in Figure~\ref{dhfig}
for $\kappa = 5$.

The string susceptibility exponent $\gamma_s$ defines the
sub-leading correction to the large-volume behavior of the 
canonical partition function: 
$\Omega_{N_3} \sim \exp \{ \mu_c N_3\} N_3^{\gamma_s -3}$.
An efficient method for determining $\gamma_s$ in
quasi-canonical simulations is provided
by the size distribution of baby universes.
A baby universe is a part of a triangulation
connected to the rest via a small ``neck'', usually
chosen as a minimal neck \cite{sbaby}.  
For a three-dimensional combinatorial triangulation a minimal neck 
consists of four triangles glued together along their
edges into a closed surface (a tetrahedra).  By cutting the
triangulation up along this boundary it is divided into
two parts; a ''mother`` of size $N_3 - B$ and a ''baby`` of size $B$.  
The size distribution of baby universes,
$b_{C_4}(B)$, is defined as all such partitions 
averaged over the ensemble.
From the asymptotic behavior of the canonical partition
function it follows that
\begin{equation}
b_{C_4}(B) \;=\; \frac{B \; \Omega_B \; (N_3 - B) \; \Omega_{N_3-B}}
 {\Omega_{N_3}} 
   \; \sim \; \left [ B(N_3 - B) \right ]^{\gamma_s - 2}
\label{baby}
\end{equation}
The distribution $b_{C_4}(B)$ can be measured with high accuracy 
in quasi-canonical simulations and the exponent $\gamma_s$ is
extracted by a fit to Eq.~(\ref{baby}).

The above definition of a baby universe also applies to 
degenerate triangulations although in this case
a minimal neck consist of only {\it two} triangles 
glued together along their three edges 
(see e.g.\ Figure~2{\it b}).  This defines a distribution
of baby universes, $b_{D_2}(B)$, separated from the
rest by two links in the dual graph.  It is also possible
to define a distribution of baby universes, $b_{D_4}(B)$,  
separated from the rest by four links in the dual graph,
analogous to the definition for combinatorial
triangulations. I have verified that
the two distributions, $b_{D_2}(B)$ and $b_{D_4}(B)$, lead
to the same estimate of $\gamma_s$.

I measured the distributions for $\kappa = 5$ and on
volumes $N_3 = 200$, 400 and 800, and estimated $\gamma_s$ 
by a fit to Eq.~(\ref{baby}).  This gave the 
following estimates: $\gamma_s = 0.47(3)$, 0.48(3) and
0.44(4), for $N_3 = 200$, 400 and 800, respectively.
This agrees reasonable with the expected value for
branched polymers, $\gamma_s = 1/2$, especially given
the small size of the triangulations used in the estimate.

\section{Discussion}

I have investigated a model of three-dimensional simplicial
gravity defined with an ensemble of degenerate triangulations
allowing distinct simplexes sharing the same set of vertexes.
The motivation for using degenerate triangulations, instead
of the combinatorial ones traditionally used in simplicial
gravity for $D > 2$, is the experience from two-dimensional
simplicial gravity which shows that using less restricted  
triangulations, i.e.\ a larger triangulations-space,
substantially reduces the finite-size effects in
numerical simulations.

As in two dimensions using a different ensemble of triangulations,
provided it yields a well defined statistical model,
does not seem to change the phase structure of the model;
I still observe a crumpled and a branched polymer phase,
separated by a discontinuous phase transition.  There is,
however, evidence of less finite-size effects in
the model, especially in the crumpled phase where the 
critical cosmological constant
approaches much faster its infinite volume value.  As a
consequence, an exponential bound on the number of
degenerate triangulations of a given volume is more easily established
than for the corresponding ensemble of
combinatorial triangulations.  
This reduction in finite-size
effects in the crumpled phase is intuitively clear; 
with less restrictions on the connectivity of the simplexes
the degenerate triangulations crumple more easily.
In addition, the discontinuous nature of the phase transition
is easily observed on triangulations of relatively
modest size.

A natural extension of the work presented in this paper is to
add a measure term \cite{measure}, or alternatively matter fields
\cite{3dmatter},
to the action Eq.~(\ref{model}) and to investigate if the phase
structure of the model is modified in the same way as has
recently been observed in simulations with combinatorial 
triangulations in both four \cite{us4d} and three 
dimensions \cite{us3d}.  
There a new phase --- a {\it crinkled} phase --- appears for
a sufficiently large negative coupling to the measure 
term, or if several matter fields are coupled to the model.
If this phase structure does indeed represent some
behavior of the continuum it is plausible, based on the
expectation of universality, that it should not depend on 
the details of the discretization 
and should also be present for degenerate 
triangulations. Thus corresponding simulations of modified
models of three-dimensional simplicial gravity using 
degenerate triangulations could serve as a confirmation on
the observed phase structure.
This work is in progress.

Although the result of my numerical investigation of
maximally degenerate three-dimensional triangulations indicated 
that this ensemble does not lead to a well defined statistical
model, I do not rule out the possibility that a less
restricted ensemble of triangulations, than considered
in this paper, could be used.  The challenge is to identify
an appropriate easing of the restriction which 
defines a consistent, exponentially bounded, canonical ensemble.  
With the potential benefits provided by an even larger ensemble
than restricted degenerate triangulations 
this is definitely worth investigating.

\vspace{20pt}
\noindent
{\bf Acknowledgments:}
I would like to thank Piotr Bialas, Sven Bilke
and Bengt Petersson for stimulating discussions.
This work was supported by the Humboldt Foundation.


\begin{thebibliography}{99}

{\small
\raggedright

\bibitem{david}
 F. David, {\it Simplicial Quantum Gravity and Random
 Lattices}, ({\tt hep-th/9303127}), Lectures given at Les
 Houches Summer School on Gravitation and Quantizations,
 Session LVII, Les Houches, France, 1992;

\bibitem{janbook}
 J.~Ambj\o rn, B.~Durhuus and T.~Jonsson, 
 {\it Quantum Geometry: A statistical field theory approach},
 (Cambridge University Press, 1997).

\bibitem{jan3d}
 J.~Ambj\o rn and S.~Varsted,
  Phys.~Lett.~{\bf B266} (1991) 285; 
  Nucl.~Phys.~{\bf B373} (1992) 557;
 D.V. Boulatov and A. Krzywicki,
   Mod.~Phys.~Lett.~{\bf A6} (1991) 3005; \\
 J.~Ambj\o rn,~D.V.~Boulatov, A.~Krzywicki and S.~Varsted,
  Phys.~Lett.~{\bf B276} (1992) 432.
  
\bibitem{sim3d}
  S.~Catterall, J.~Kogut and R.~Renken,
   Phys.~Lett.~{\bf B342} (1995) 53;
   Nucl.~Phys.~{\bf B523} (1998) 553.
   
\bibitem{jap3d}
 T.~Hotta, T.~Izubuchi and J.~Nishimura,
  Nucl.~Phys.~{\bf 63} (Proc.~Suppl) (1998) 63;
  {\it Multicanonical simulation of 3-d dynamical triangulation
    model and a new phase structure}, ({\tt hep-lat/9802021}).
    
\bibitem{jab3d2}
  H.~Hagura, N.~Tsuda and T.~Yukawa,
   Phys.~Lett.~{\bf B418} (1998) 273; \\
  H.S.Egawa, N.Tsuda and T.~Yukawa, {\it Common structures in
   simplicial quantum gravity}, ({\tt hep-lat/9802010}).

\bibitem{mat2}
  E.~Brezin, C.~Itzykson, G.~Parisi and J.B.~Zuber,
   Commun.~Math.~Phys.~{\bf 59} (1978) 35.  
    
\bibitem{medeg}
 J.~Ambj\o rn, G.~Thorleifsson and M.~Wexler,
  Nucl.~Phys.~{\bf B439}(1995) 187.
  
\bibitem{minimal}
  M.~Bowick, S.~Catterall and G.~Thorleifsson,
   Phys.~Lett.~{\bf B391}(1997) 305;
   Nucl.~Phys.~{\bf 53} (Proc.Suppl) (1997) 753; \\
  V.A.~Kazakov, M.~Staudacher and T.~Wynter,
   Nucl.~Phys.~{\bf B471} (1996) 309.   
   
\bibitem{move}
 N.~Godfrey and M.~Gross, 
  Phys.~Rev.~{\bf D43} (1991) R1749; \\
 M.~Agishtein and A.~Migdal, 
  Mod.~Phys.~Let.~{\bf A6} (1991) 1863; \\  
 M.~Gross and S.~Varsted,
  Nucl.~Phys.~{\bf B378} (1992) 367.

\bibitem{simprog}
  S.~Catterall, Comput.~Phys.~Commun.~{\bf 87} (1995) 409.
  
\bibitem{sing4d} 
 T.~Hotta, T.~Izubuchi and J.~Nishimura,
  Prog. Theor. Phys. {\bf 94} (1995) 263; \\
 S.~Catterall, G.~Thorleifsson, J.~Kogut and R.~Renken,
  Nucl.~Phys.~{\bf B468} (1996) 263; \\
 P.~Bialas, Z.~Burda, B.~Petersson and J.~Tabaczek,
  Nucl.~Phys.~{\bf B495} (1997) 463; \\
 S.~Catterall, J.~Kogut and R.~Renken,
  Phys.~Lett.~{\bf B416} (1998) 274.

\bibitem{binder}
 K.~Binder, in {\it The Monte Carlo method in condensed matter
  physics}, ed.~K.~Binder, 2nd edition, Springe-Verlag 1992.

\bibitem{scaling}
 S.~Catterall, G.~Thorleifsson, M.~Bowick and V.~John,
  Phys.~Lett.~{\bf B354} (1995) 56;
 J.~Ambj\o rn, J.~Jurkiewicz and Y.~Watabiki,
  Nucl.~Phys.~{\bf B454} (1995) 313.

\bibitem{sbaby}
 S.~Jain and S.D.~Mathur, Phys.~Lett.\ {\bf B286} (1992) 239; \\
 J.~Ambj\o rn, S.~Jain and G.~Thorleifsson,
  Phys.~Lett.~{\bf B307} (1993) 34; \\
 J.~Ambj\o rn and G.~Thorleifsson, 
  Phys.~Lett.~{\bf B323} 7(1994) 7.

\bibitem{measure}
 B.~Brugmann and E.~Marinari,
  Phys.~Rev.~Lett.~ {\bf 70} (1993) 1908.

\bibitem{3dmatter}
 R.L.~Renken, S.M.~Catterall and J.~Kogut,
  Nucl.~Phys.~{\bf B422} (1994) 677; \\
 J.~Ambj\o rn, J.~Jurkiewicz, S.~Bilke, Z.~Burda and B.~Petersson,
  Mod.~Phys.~Lett.~{\bf A9} (1994) 2527; \\
 J.~Ambjorn, Z.~Burda, J.~Jurkiewicz and C.F.~Kristjansen,
  Phys.~Lett.~{\bf B297} (1992) 253.
 
\bibitem{us4d}
 S.~Bilke, Z.~Burda, A.~Krzywicki, B.~Petersson, J.~Tabaczek
  and G.~Thorleifsson, Phys.~Lett.~{\bf B418} (1998) 266;
  {\it 4D simplicial quantum gravity: Matter fields and the
  corresponding effective action}, ({\tt hep-lat/9804011}).
  
\bibitem{us3d}
 B.~Petersson, O.~Prauss, G.~Thorleifsson and T.~Westheider,
  in preperation.
  
}

\end{thebibliography}
\end{document}